\documentclass[a4paper,showkeys,floatfix,aps,prl,reprint,groupedaddress]{revtex4-1}
\usepackage{amsmath,amssymb}
\usepackage{graphics,graphicx}
\usepackage{dcolumn,bm}
\usepackage{psfrag}
\usepackage{soul}
\usepackage{hyperref}
\usepackage{float}
\hypersetup{colorlinks=true, citecolor=blue, urlcolor=blue, linkcolor=blue}

\begin{document}

\title{Inequality indices for heterogeneous systems: a tool for failure prediction}

\author{Tarun Ram Kanuri\,$^{1,*}$, Subhadeep Roy\,$^{2}$, Soumyajyoti Biswas\,$^{1}$}
\affiliation{$^{1}$ Department of Physics, SRM University - AP, Andhra Pradesh 522240, India.\\
  $^{2}$ Department of Physics, BITS Pilani Hyderabad Campus, Secunderabad, Telangana 500078, India.}
\email{tarunram_kanuri@srmap.edu.in}
\email{subhadeep.r@hyderabad.bits-pilani.ac.in}
\email{soumyajyoti.b@srmap.edu.in}

\date{\today}

\begin{abstract}

    We have numerically studied a mean-field fiber bundle model of fracture at a non-zero temperature and acted by a constant external tensile stress. The individual fibers fail (local damage) due to creep-like dynamics that lead up to a catastrophic breakdown (global failure). We quantify the variations in sizes of the resulting avalanches by calculating the Lorenz function and two inequality indices -- Gini ($g$) and Kolkata ($k$) indices -- derived from the Lorenz function. We show that the two indices cross just prior to the failure point when the dynamics goes through intermittent avalanches. For a continuous failure dynamics (finite numbers of fibers breaking at each time step), the crossing does not happen. However, in that phase, the usual prediction method i.e., linear relation between the time of minimum strain-rate and failure time, holds. The boundary between continuous and intermittent dynamics is very close to the boundary between crossing and non-crossing of the two indices in the temperature-stress phase space.       
\end{abstract}


\maketitle

\section{Introduction}

Failure process in disordered solids \cite{hr90} depends on environment, such as pressure, temperature; driving condition such as applied stress, strain rate; as well as structural or material properties such as defects, micro-cracks, plasticity, etc. One important scenario in this regard is that a material can break with time even if a constant stress, less than the critical value is applied on it - a phenomena widely know as the \textit{creep} \cite{nhgs05,nhgs05a,scirep} in the material science community. Especially, the creep process shows a remarkable change as the temperature is tuned due to the temperature-dependent dislocation motion, which becomes even easier when the thermal fluctuation is higher. A detailed study of the creep failure giving an estimation of creep time is very crucial for damage control regarding foundations like bridges, building blocks and other constructions where the material in constantly under an external load. Apart from the obvious effect of temperature on the creep process, structural randomness and heterogeneity also plays an important role as the creep takes place with the accumulation of plastic strain and damages with time. The difficultly of handling the creep in heterogeneous solids leads to other approaches, i.e, sub-critical crack growth dynamics \cite{m00} and phenomenological Voight's model for precursory strain \cite{v89}, where a time-dependent strain similar to experiments are observed assuming a heterogeneous feedback system. 

For the present work, we will adopt a fiber bundle model \cite{hansen15_book}, which is a statistical model lead by threshold activated dynamics \cite{rc85}. The model is simple yet very powerful to simulate failure dynamics under different scenarios including a time-dependent creep process. Some studies considered fluctuation in applied stress in form of thermal fluctuation \cite{cgs01}. In some studies, the failure events are assumed to be probabilistic and not deterministic in presence of a finite temperature \cite{pcc13}. Other studies involve complex rheological properties that depend on specific rheological properties of fibers. For instance, the Kelvin-Voigt rheology \cite{hkh02,kmhh03}, a nonlocal rheology \cite{j11} or a damage rheology for each fiber \cite{dk13}. Here, we consider a ELS fiber bundle model but at a constant temperature instead. A non-zero temperature will add an extra source of fluctuation, the thermal one, in addition to the fluctuation in local strength we already have among the fibers. Though FBM has shown athermal creep \cite{rh18} when the applied stress is closer to the critical value, a finite temperature makes the rupture events probabilistic, making the failure process reminiscent of a thermal creep failure where a material can break under any stress no matter how low it is. Such a probabilistic approach has been reported in the in the literature exploring a time-dependent relaxation dynamics \cite{rh20}. For the present paper, we want to combine the probabilistic approach in FBM to mimic creep with the inequality indices obtain from a generic form of the Lorenz function \cite{lorenz05} to predict an upcoming catastrophic failure. 

\begin{figure*}[t]
\centering
\includegraphics[width=6.0cm, keepaspectratio]{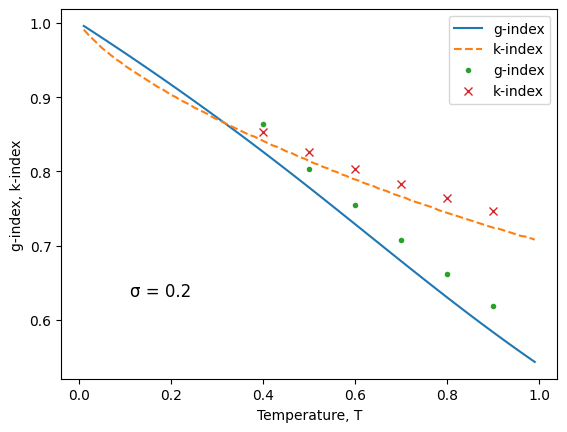}  \includegraphics[width=6.0cm, keepaspectratio]{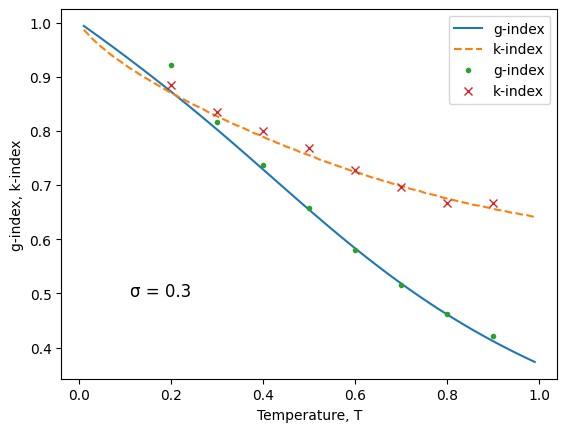} \\
\includegraphics[width=6.0cm, keepaspectratio]{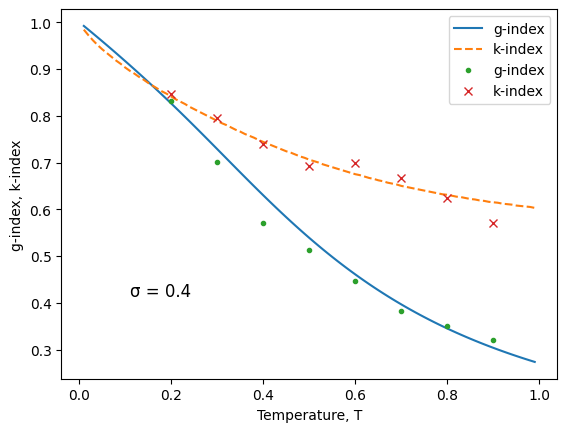}  \includegraphics[width=6.0cm, keepaspectratio]{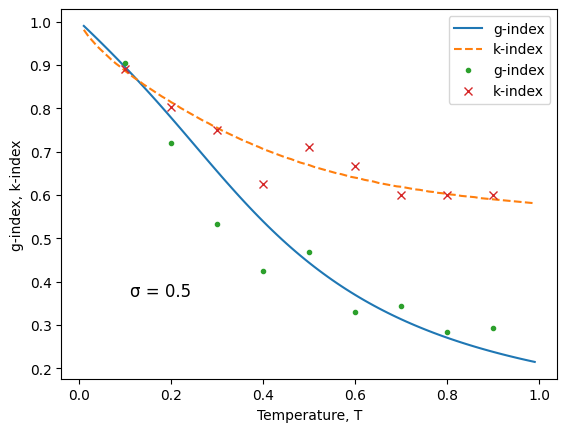}
\caption{The figure shows the variation of Gini index $g$ and Kolkata index $k$ as a function of temperature $T$ and $\sigma=0.2$, 0.3, 0.4, and 0.5. $g$-index and $k$-index compete with each other and becomes equal at 0.87. The threshold distribution is a delta function at 1.0.}
\label{fig1}
\end{figure*}

Lorenz function $L(p)$ has been introduced in the context of wealth distribution and represents cumulative wealth fraction possessed by the $p$ fraction of the population. For poorest and richest the $L(p)$ function takes the form $0$ and $1$ respectively; $L(p)=p$ being the line of equality, showing equal distribution of wealth. The two most popular equality index in this context are as follows: (i) {Gini index (g) \cite{gini21}: $g=0$ corresponds to perfect equality, $g=0$ for extreme inequality and (ii) Kolkata index (k) \cite{ghosh14}: $k=1/2$ corresponds to perfect equality, $g=1$ for extreme inequality. We get $g=k$, as per the law old Pareto \cite{pareto71} at an extremely competitive situation between equality and inequality.  In the present paper, we have used the same framework by replacing wealth by the avalanches during a failure process, where an avalanche of size $s$ can occur with a probability $p$.  

\section{Model Description}

Numerically, we have studied a mean-field fiber bundle model  (FBM) \cite{hansen15_book} at a non-zero temperature with an external load applied on it. A conventional FBM consists of a bunch (say $L$) of vertical fibers attached between two parallel bars (in case of 1d) or plates (in case of 2d), which are pulled apart with an external force $F$ creating a stress $\sigma (=F/L)$ on each fiber. The disorder is introduced in the model through the fluctuation in strength from fiber to fiber. This is done by choosing the threshold stress of each fiber randomly from a distribution. For the present work, we have chosen both the Uniform and the Weibull distribution given as follows: 
\begin{equation}\label{eq3}
    \rho(x)=
\begin{cases}
\displaystyle\frac{1}{2\Delta} \ \text{for} \ 1-\Delta \le x \le 1+\Delta, \\
0 \ \text{elsewhere}
\end{cases}
\end{equation}
\begin{equation}\label{eq3}
    \rho(x)= \displaystyle\frac{k}{\lambda}\left(\displaystyle\frac{x}{\lambda}\right)^{k-1}e^{(-x/\lambda)^k}
\end{equation}
In case of the Uniform distribution, the span $\Delta$ of the distribution is the measure of disorder while for Weibull, the disorder strength is tuned by varying the shape ($k$) and scale parameter ($\lambda$). 

\begin{figure}[ht]
\centering
\includegraphics[width=7.0cm, keepaspectratio]{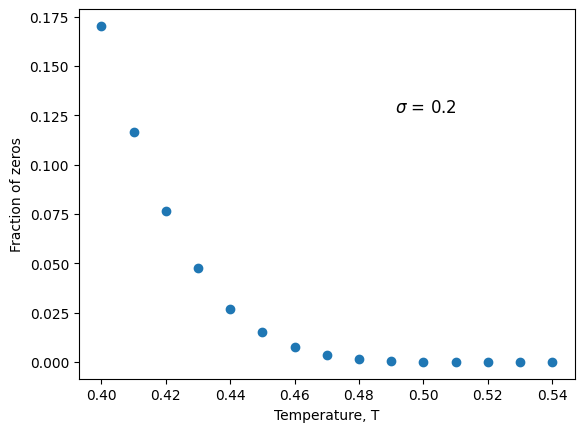}  
\caption{The figure shows the number of zeros versus the temperature for $\sigma=0.2$ and a delta-distribution. As $T$ increases, we observe a decrease in the number of zeros and finally they saturates at a low value at high temperature.}
\label{fig2}
\end{figure}
\begin{figure*}[t]
\centering
\includegraphics[width=6.0cm, keepaspectratio]{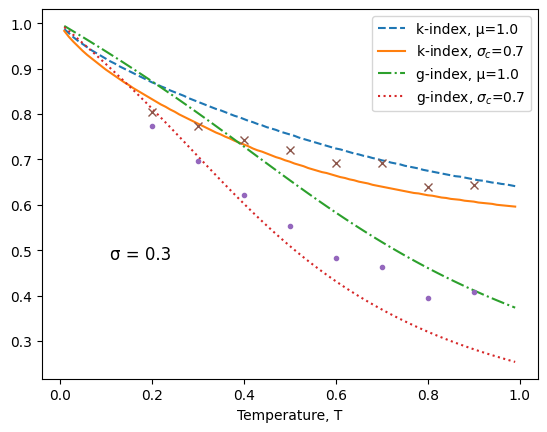}  \includegraphics[width=6.0cm, keepaspectratio]{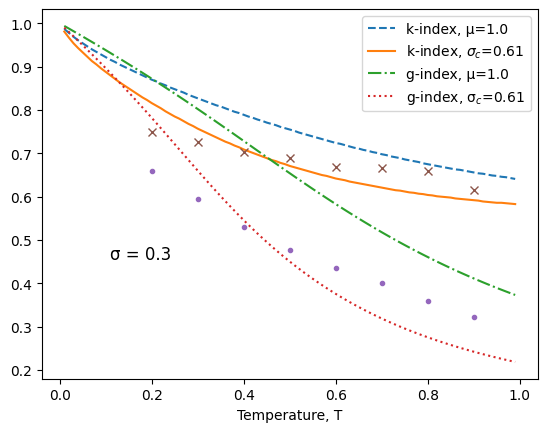} \\
\includegraphics[width=6.0cm, keepaspectratio]{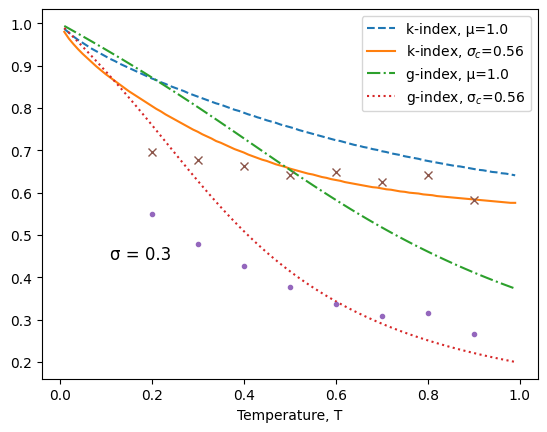}  \includegraphics[width=6.0cm, keepaspectratio]{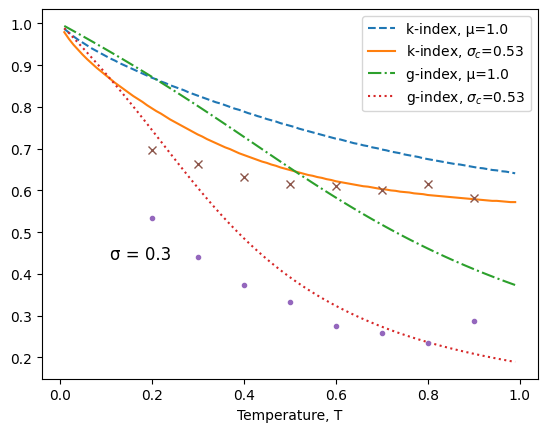}
\caption{The figure shows the variation of Gini index $g$ and Kolkata index $k$ as a function of temperature $T$ and $\sigma=0.3$. As a threshold distribution, we have used a uniform distribution spanning from $1-\Delta$ to $1+\Delta$. The disorder strength $\Delta$ is set to be 0.3, 0.4, 0.5, and 0.6. $\mu$ and $\sigma_c$ are respectively the mean strength and the critical stress for quasi-static lading. $g$-index and $k$-index compete with each other and becomes equal at 0.87.}
\label{fig3}
\end{figure*}

Once the external stress crosses the threshold value for a certain fiber, that fiber breaks irreversibly and the stress carried by the broken fiber is redistributed among the rest of the model. In this paper, we adopted a equal-load sharing (ELS) scheme \cite{p26} where the stress after each rupture is redistributed equally among all surviving fibers. This is also the mean-field limit of the model. Apart from ELS scheme there are other redistribution schemes that has been explored in the context of FBM: (i) local load sharing (LLS) scheme \cite{p79} where the stress is redistributed among the nearest neighbours only, (ii) mixed mode redistribution \cite{pch05} where the stress redistribution depends on how far a fiber is from the broken one, etc. 

As mentioned in the introduction, inclusion of temperature makes the rupture events probabilistic depending on the temperature as well as applied stress: 
\begin{equation}\label{eq3}
P(\sigma,T)=\frac{\sigma}{\sigma_c}exp\left[-\frac{1}{T}\left(\frac{\sigma_c}{\sigma}-1\right)\right]
\end{equation} 
where, $\sigma$ and $\sigma_c$ are respectively the local stress and threshold stress of a fiber. $\sigma_c$ is either set to be either 1 (equal for all fibers) or chosen from a random distribution (say Uniform or Weibull). Due to such a probabilistic approach, a fiber can break even when $\sigma<\sigma_c$. The relaxation dynamics of the model goes as follows:
\begin{enumerate}
    \item At each time we chose a probability $P_i(\sigma,T)$ for a fiber $i$ and compare it with a random number $r$ chosen with uniform probability between 0 and 1. 
    \item A fiber breaks irreversibly if $P_i(\sigma,T)>r$ and the stress of the fiber is redistributed equally among all other surviving fibers increasing the value of $\sigma$ and hence the probability $P$ (see Eq.\ref{eq3}).
    \item In the next time step we chose another $P$ depending on temperature $T$ and the updated local stress $\sigma$ and repeat the previous step.
    \item This process goes on until all fibers are broken suggesting global failure. 
\end{enumerate}
We want to stress the following point here. For $T=0$, $P(\sigma,T)$ in equation \ref{eq3} becomes zero suggesting no fibers will break unless $\sigma \approx \sigma_c$.

\section{Analysis}

Consider an equal load sharing FBM, where all fibers have the same failure threshold $\sigma_c$. There is an external noise $T$. Now, under a load $\sigma$, a fiber will break if $\sigma \ge \sigma_c$. But even when $\sigma< \sigma_c$, the
fiber can break with a probability given by Eq.\ref{eq3}. 

Now, if a time step $t$ is defined as one step of load redistribution, then the fraction of surviving fibers
is given by
\begin{equation}\label{eq4}
U(t,\sigma,\sigma_c,T)=\frac{\sigma T}{\sigma_c}ln\left[\frac{\tau -t}{Texp(-1/T)}+1\right],
\end{equation}\label{eq5}
where $\tau$ is the failure time of the system, given by
\begin{equation}\label{eq6}
\tau=T exp\left(-\frac{1}{T}\right)\left[exp\left(\frac{\sigma_c}{\sigma T}\right)-1\right].
\end{equation}\label{eq7}
The magnitude of an avalanche, defined as the number of fibers failing in one time step, is then 
\begin{align}\label{eq8}
S(t)=\left|\frac{dU}{dt}\right|=\frac{\sigma T}{\sigma_c}\frac{1}{(\tau-t)+Texp(-1/T)}.
\end{align}
The Lorenz function is then
\begin{equation}\label{eq9}
L(p)=\frac{\int\limits_0^{p\tau}S(t)dt}{\int\limits_0^{\tau}S(t)dt}.
\end{equation}
This on simplification gives,
\begin{equation}\label{eq10}
L(p)=\frac{ln\left|T exp(-1/T)+\tau\right| - ln\left|T exp(-1/T)+(1-p)\tau\right|}{ln\left|T exp(-1/T)+\tau\right| - ln\left|T exp(-1/T)\right|}
\end{equation}

The value of the Gini index $g$ at the failure point can be calculated from
\begin{equation}\label{eq11}
g_f=1-2\int\limits_0^1L(p)dp.
\end{equation}
This on simplification gives
\begin{equation}\label{eq13}
g=1+2\left[exp\left(1-\frac{\sigma_c}{\sigma T}\right)-\frac{1}{ln\left(1+exp\left(\frac{\sigma_c}{\sigma T}\right)-1\right)}\right].
\end{equation}
The value of the Kolkata index ($k$) at the failure point can be obtained from
\begin{equation}\label{eq14}
1-k=L(k),
\end{equation}
which needs to be numerically solved.

\section{Numerical Results}

In this section, we will discuss the numerical results that will build towards the discussion as well as the main finding of the present article. As already mentioned in the {\it Model Description} section, the numerical simulation is done for a mean field fiber bundle model with system size $10^5$ and $10^4$ realizations. For the first part of the numerical results, we have considered that all fibers have the same breaking threshold equals to unity while in the later part Uniform and Weibull distributions are considered. 

Figure \ref{fig1} the variation of the {\it Gini index} and {\it Kolkata index} as we vary the external temperature, which tunes the strength of the thermal fluctuation. The distribution of the threshold values are chosen from a $delta$ function with it's peak at 0.5. Due to the choice of the distribution, the fluctuation among the local threshold values is eliminated. Though, this does not eliminate the thermal fluctuation (equation \ref{eq3}) and the rupture events are still probabilistic and not deterministic. The model is observed under four different external stress values: 0.2, 0.3, 0.4 and 0.5. The points corresponds to the numerical results for a $\delta$-distribution at 0.5. The dotted lines on the other hand is the comparison of the numerical results with our analytically observed {\it g-index} and {\it k-index}. The behavior can be divided in two different regions:
\begin{itemize}
\item Region I: This region is observed at a higher temperature where the thermal fluctuation is sufficiently large. We observe the Gini index here to be less than the $k$-index. Such behavior is evident through both analytical results as well as numerical data. 
\item Region II: This region takes place where the temperature is low and hence the thermal fluctuation is much less. The fiber rupture events are almost deterministic here since the probabilistic element is negligible due to this very small temperature. Here the $g-$index overcomes $k-$index and becomes larger. 
\end{itemize}


\begin{figure}[ht]
\centering
\includegraphics[width=7.0cm, keepaspectratio]{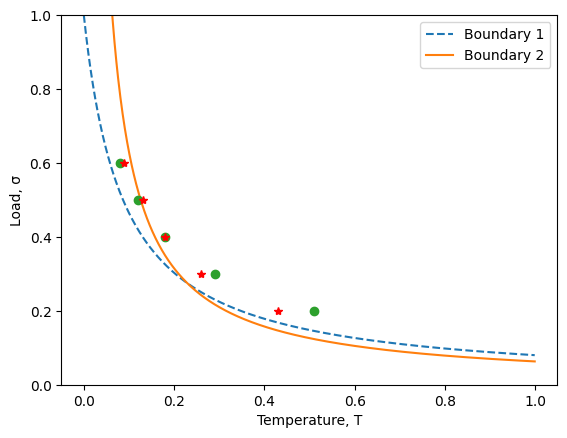}  
\caption{Boundaries showing the separation of {\it Region I} and {\it Region II}. Boundary 1 is drawn from the direct comparison of $g-$index and $k-$index. Boundary 2 is drawn where the number of zeroes disappears from the system dynamics.}
\label{fig5}
\end{figure}
\begin{figure*}[ht]
\centering
\includegraphics[width=6.0cm, keepaspectratio]{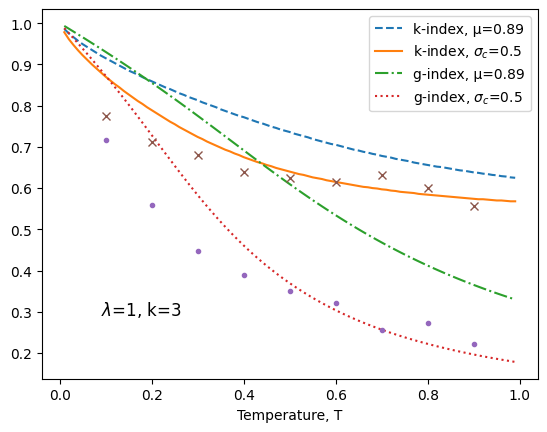}  \includegraphics[width=6.0cm, keepaspectratio]{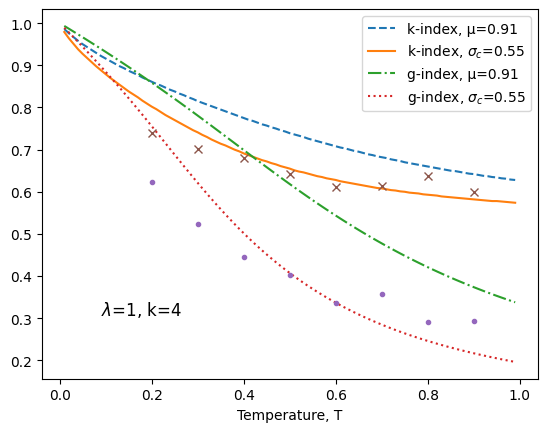} \\
\includegraphics[width=6.0cm, keepaspectratio]{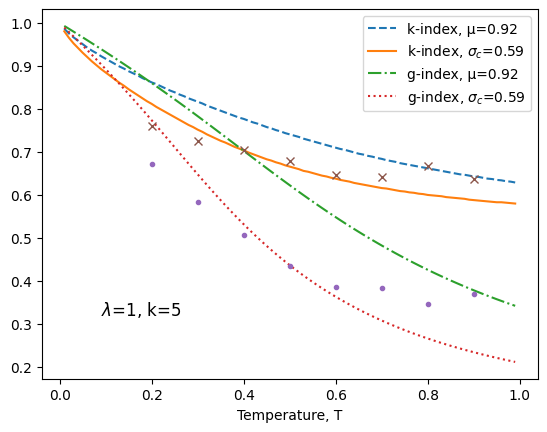}  \includegraphics[width=6.0cm, keepaspectratio]{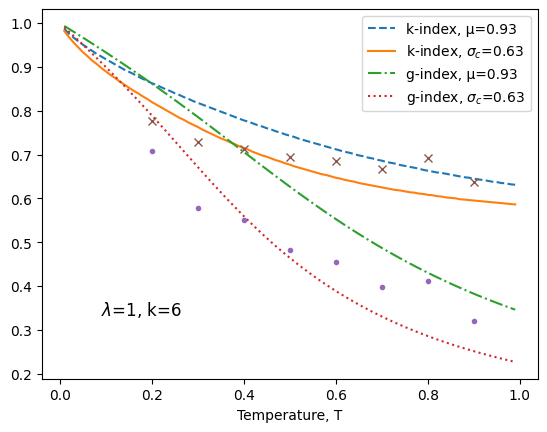}
\caption{The figure shows the variation of Gini index $g$ and Kolkata index $k$ as a function of temperature $T$ and $\sigma=0.3$. As a threshold distribution, we have used a Weibull distribution of scale parameter $\lambda$ and shape parameter $k$. $\mu$ and $\sigma_c$ are respectively the mean and the critical stress for quasi-static lading. $g$-index and $k$-index compete with each other and becomes equal at 0.87. .}
\label{fig6}
\end{figure*}

As $\sigma$ increases, we observe the crossover between region I and region II starts to take place at a lower temperature. Lets call this point $T^{\ast}$. We observe $T^{\ast}\approx0.38$ for $\sigma=0.2$ while for $\sigma=0.5$, we get $T^{\ast}\approx0.15$. Though $T^{\ast}$ changes with $\sigma$, the crossover takes place for $g=k=0.87$ irrespective of the value of external stress. This universal value has been observed earlier in the context of other statistical systems as well, like Bak-Tang-Wiesenfeld (BTW) \cite{bcw87}, Manna model \cite{m91}, Edwards-Wilkinson (EW) \cite{ew82}, interface dynamics based on fiber bundle model \cite{bc13}, etc.      

Next we turn to find another way to understand the crossover point when $g$ becomes more than $k$. Since the dynamics of the model is lead by the thermal fluctuation, the probability that a certain fiber will break decreases as we go to lower temperature or lower applied stress. Which means, at lower $T$ and $\sigma$, there will be many time steps in-between where no rupture events happen. We call this the {\it zeroes} of the model evolution. Figure \ref{fig2} shows the number of {\it zeroes} as a function of temperature when the external stress is kept at $\sigma=0.2$. For low $T$ there are many {\it zeroes} observed while it decreases and becomes gradually 0 after a certain temperature (say $T^{\prime}$).  
Figure \ref{fig3} shows the same result as figure \ref{fig1} but for a uniform distribution spanning between $1-\Delta$ to $1+\Delta$, $\Delta$ being the strength of disorder. Since the analysis becomes complicated due the fluctuation of strength from fiber to fiber (since its a distribution now instead of a delta-function). To chose a certain $\sigma_c$ in equation \ref{eq13} and \ref{eq14}, we adopted the following assumptions: (i) since its a uniform distribution, we can assume the mean ($\mu$) to be the critical value; (ii) we can drive the system in a quasi-static manner at zero temperature and found out the average critical stress and use it. 

In figure \ref{fig3}, we have kept the same external stress $\sigma=0.3$ but changed the width of the uniform distribution, and hence the strength of disorder. We observe a far better fit for $g-$index (equation \ref{eq13}) and $k-$index (equation \ref{eq14}) when we consider $\sigma_c$ to be the critical stress for quasi-static loading rather than the mean strength. Also, quality of such fit decays as we go to higher disorder strength (high $\Delta$). 

We want to argue here that, $T^{\ast}$ where $g$ becomes greater than $k$, and $T^{\prime}$ where number of {\it zeroes} itself becomes 0 has the same values. We will discuss this in details next. Figure \ref{fig5} shows this boundary between Region I and Region II, independently calculated from (i) intermittence behaviour of the system (study of fraction of {\it zeroes}) and (ii) the interplay of $g$ and $k$, respectively. The points corresponds to the numerical results found from direct simulation. The solid lines, on the other hand, are drawn from the following analytical approaches: (i) {\bf Boundary 1:} This shows the ($\sigma, T$) coordinates beyond which there are no {\it zeroes} in avalanches, suggesting that at least one fiber will break at each time step. This is done by using the analytical results from the paper by Pradhan et al \cite{pcc13}. (ii) {\bf Boundary 2:} This boundary consists of the coordinates below which $g$ overcomes $k$. This boundary is obtained by numerically solving the equations \ref{eq13} and \ref{eq14}. The bottom line this these boundaries, {\it Boundary 1} or {\it Boundary 2}, separates two regions I and II where the method of failure prediction is vastly different. {\it Region I}, is the area above the boundary which is accessible by increasing either the temperature of the external stress. In this region, we observe fiber ruptures at each and every time step. The usual linear relation between the time of minimum strain-rate and failure time holds here and the prediction of upcoming failure can be done through this relation. On the other hand, for low $T$ and $\sigma$, we obtain {\it Region II}, where intermittent behavior is observe during the rupture events and above linear relationship does not hold good. However, in this region the failure point can be predicted from the the competition of $g$ and $k$ index and the Gini index is observed to show higher value than $k$-index just before the catastrophic failure.     

Figure \ref{fig6} finally shows the comparison of the analytical and numerical behavior of $g$ and $k$ indices for a Weibull distribution of shape parameter $k$ and scale parameter $\lambda$. This shows the same fact that the fitting is better by considering the critical stress for a quasi-brittle loading rather than the mean of the distribution. This strengthen our claim regarding the universality of our results irrespective of the choice of the threshold distribution.  
 

\section{Discussion}
Under a constant load, which is less than the critical load of a disordered solid, there are creep induced failures that lead to the eventual collapse of the system in presence of external noise e.g., temperature. The fiber bundle model of fracture reproduces essential features of such failure through a simple mean-field dynamics. 

Here we have shown that the resulting avalanches of the creep failure can be quantified in terms of their variations in sizes. Particularly, a Lorenz function for the avalanches under a constant load and temperature can be analytically calculated for a delta-function threshold distribution with it's peak at 1. Consequently, the Gini ($g$) and Kolkata ($k$) indices can also be derived (see Eq. (\ref{eq13}) and (\ref{eq14})). Curiously, the time variations of $g$ and $k$ (cumulative measured of inequality until that time) cross each other just prior to the failure for sufficiently low values of external load and thermal noise that ensure an intermitten avalanche dynamics. For large load and/or thermal noise, the avalanche dynamics become continuous, where such crossing do not happen. In that region, however, the linear relation between the time of strain-rate minimum and failure time is well maintained. 

For wides distributions of failure thresholds, for example Weibull or uniform, the observations are qualitatively similar (see Figs. \ref{fig3}, \ref{fig6}). However, the analytical expression for such cases could be done by approximating the threshold distribution by its smallest value (for narrow threshold distributions) or the average value (for wide distributions). 

In conclusion, for intermittent creep rupture dynamics, the variations in sizes of the resulting avalanches grow with time and provide a very reliable precursory signal for the imminent failure in a wide range of the temperature-stress phase space, where traditional precursory indications do not work.


\end{document}